\documentclass[11pt]{article}

\usepackage{graphicx}
\usepackage{dcolumn}
\usepackage{bm}

\begin{document}

\centerline{\bf  THE MICROTUBULE TRANSISTOR}

\bigskip

\centerline{H.C. Rosu\footnote{hcr@ipicyt.edu.mx; file MTT.tex}}

\begin{center}
Potosinian Institute of Science and Technology (IPICyT)\\ Apdo
Postal 3-74 Tangamanga, 78231 San Luis Potos\'{\i}, Mexico
\end{center}

\date{\today}  


\begin{abstract}
I point out the similarity between the microtubule (MT) experiment
reported by Priel {\em et al} \cite{priel} and the ZnO nanowire
experiment of Wang {\em et al} \cite{xwang}.  It is quite possible
that MTs are similar to a piezoelectric field effect transistor
(PE-FET) for which the role of the control gate electrode is played
by the piezo-induced electric field across the width of the MT walls
and their elastic bending features.
\end{abstract}




 \bigskip

 \thispagestyle{empty}




\section{The MT transistor-like behavior}

A very interesting experiment has been published recently by Priel
and collaborators \cite{priel}. Isolated MTs were visualized and
identified under {\em phase contrast microscopy} and connected to
the tips of two {\em
patch-clamp amplifiers }
such that electrical stimulation could be applied to one of them,
whereas the electric signal could be collected with the other one.

\medskip

MTs were electrically stimulated by applying 5-10 ms input voltage
pulses with amplitudes in the range of $\pm \,200$ mV. The resulting
electrical signals were obtained at the opposite end of an MT, 20-50
$\mu$m away, with a pipette also connected to a patch amplifier,
which was kept ``floating" at 0 mV.

\medskip

Two remarkable findings are reported: \\

\noindent First, coupling of the stimulus pipette to an MT increased
the overall conductance of the pipette by $> 300$ \%.

\medskip

\noindent Second, the signals reaching the collection site were in
about one third of the cases
higher than those obtained in free solution. 

\medskip

\noindent Moreover, the MT amplified both the electrical pulse
injected at the
stimulus site and the collection site as well. 

\medskip

Thus the overall conclusion is that MTs improve electrical
connectivity between two locations in saline solution.

\medskip

The currents measured at the collection site were linearly dependent
on the stimulus pipette input voltage, indicating a strictly inverse
ohmic response, i.e., linear amplification. 
The MT conductances reached up to 9 nS, much higher than that
expected from
channel conductances (5-200 pS). 


The authors claim that their findings demonstrate that electrical
amplification by MTs is equivalent to the polymer's ability to act
as a biomolecular transistor. 
In a minimalistic way, both a constant electrical polarization and
an active component are required.

\medskip

This experiment supports previous molecular dynamics simulation of
tubulin structure \cite{12,13} indicating a strong negative surface
charge distribution of the order of 20 electrons per monomer,
distributed more on the outer surface than in the inner core with
ratio of $\sim$ 2:1. They also point to a constant (permanent)
electric polarization, which follows localized Nernst potentials
arising from asymmetries in the ionic distributions between the
intra- and extra-MT environments. This polarization is modulated by
electrical stimulation such that the forward-reverse biased
junctions of an intramolecular transistor creates a proper
MT-adjacent ionic cloud environment, which allows amplification of
of axially transferred signals. The proposed model implies that
intrinsic semiconductive like properties of the structured tubulin
dimers are such that an effective transistor is being formed whose
gating ability to modulate localized charges may help amplify axial
ionic movements.

\section{Are MTs piezoelectric field effect transistors ?}

We notice here that the experiment of Ariel {\em et al.} is very
similar to a nanotechnology experiment with a new type of FET with a
zinc oxide nanowire between two electrodes \cite{xwang}. The
electric field created by piezoelectricity across the bent nanowire
serves as the gate for controlling the electric current flowing
through the nanowire, thus it can be
tuned on/off by applying a mechanical force.\\
The experimental setup consists of a 370 nm wide, 100 $\mu$m long
zinc oxide nanowire across a tungsten needle tip and a silicon
substrate
covered with silver paint.\\
In order to test the response of the nanowire to mechanical stress,
sequential measurements were made in which the two electrodes were
approached to each other, causing the nanowire to bend. The symmetry
of the \emph{I-V} curves indicated good ohmic contacts. The current
was found to drop significantly with the increase of bending,
indicating the decreased conductance with the increased strain.
When a semiconductor crystal is under strain, the change in
electrical conductance is referred to as the piezoresistance effect,
which is usually caused by a change in band gap width as a result of
strained lattice. Some models have been proposed for homogeneous
cases, however the bending of a nanowire is by no means homogeneous.
Instead, the sample is bent so that the inner arc surface of the
nanowire is compressed ($\epsilon=\delta l / l < 0 $), and the outer
arc surface is stretched ($\epsilon > 0$), and the area close to the
center of the nanowire is strain free. Therefore, the total
piezoresistance of the nanowire can be obtained by an integration
across the nanowire cross section and its length of the usual
piezoresistance effect given by
\begin{equation}\label{nw1}
\frac{\delta \rho}{\rho}=\pi \epsilon~,
\end{equation}
where $\rho$ is the resistance, $l$ is the original length, and $\pi$ is
the piezoresistance coefficient.\\

It is not trivial to explain the observed increased resistance of
the ZnO nanowire after bending. Wang and collaborators proposed the
following explanation.  When the piezopotential appears across the
bent nanowire, some free electrons in the $n$-type ZnO nanowire may
be trapped at the positive side surface (outer arc surface) and
become non-movable charges, thus lowering the effective carrier
density in the nanowire. On the other hand, the negative potential
remains unchanged. Hence, the {\em piezo-induced electric field} is
retained across the width of the nanowire. The free electrons will
be repulsed away by the negative potential and leave a charge
depletion zone around the compressed side. Consequently, the width
of the conducting channel in the nanowire becomes smaller and
smaller while the depletion region becomes
larger and larger with the increase of the nanowire bending. \\
An almost linear relationship between the bending curvature and the
conductance was found at small bending regions. The authors have
derived the following relationship for a tranverse force and the
bending shape of the nanowire:
\begin{equation}\label{nw2}
F_{y}=\frac{3YI}{L^3}y_m~,
\end{equation}
where $F_y$ is the transverse force, $Y$ is the Young modulus, $I$
is the momentum of inertia, $L$ is the total length, and $y_m$ is
the maximum bending deflection of the nanowire (usually measured
with scanning electron microscopy techniques). From this, the
authors have concluded that for this particular nanowire the
decrease of the conductance was quasi-linear for up to 17 nN, at
which the tendency of the conductance was reduced.

\section{Conclusion}

The similarity between the two experiments can be employed to settle
a more definite transistor model for the reported amplifying
electric properties of microtubules and even for thinking of using
them in future biological force sensing devices at the single or
bundle level.


\end{document}